\documentstyle[prl,aps,epsf]{revtex}

\begin{document}

\title{
{\normalsize\hfill MPI-PhT/98-53}\\
{\normalsize\hfill July 1998}\\
The Gamma-Ray Halo from Dark Matter Annihilations}

\author{Paolo Gondolo}
\address{Max Planck Institut f\"ur Physik, F\"ohringer Ring 6, 80805 Munich, 
Germany}

\twocolumn[\hsize\textwidth\columnwidth\hsize\csname
@twocolumnfalse\endcsname

\maketitle

\begin{abstract}
\widetext
\noindent A sophisticated analysis of EGRET data has found evidence for
gamma-ray emission from the galactic halo.  I entertain the possibility that
part of the EGRET signal is due to WIMP annihilations in the halo.  I show that
a viable candidate with the required properties exists in a model with an
extended Higgs sector. The candidate has a mass of 2--4 GeV, a relic density
$\Omega \sim 0.1$ (for a Hubble constant of 60 km/s/Mpc), and a scattering
cross section off nucleons in the range $10^{-5}$--$10^{-1}$ pb. The model
satisfies present observational and experimental constraints, and makes strict
predictions on the gamma-ray spectrum of the halo emission.
\end{abstract}

\vskip 2pc]

\narrowtext

Dixon et al.~\cite{Dixon98}, in a sophisticated analysis of the EGRET gamma-ray
data, have found evidence for gamma-ray emission from the galactic halo.
Filtering the data with a wavelet expansion, they have produced a map of the
intensity distribution of the residual gamma-ray emission after subtraction of
an isotropic extragalactic component and of expected contributions from cosmic
ray interactions with the interstellar gas and from inverse Compton of ambient
photons by cosmic ray electrons. Besides a few ``point'' sources, they find an
excess in the central region extending somewhat North of the galactic plane,
and a weaker emission from regions in the galactic halo. They mention an
astrophysical interpretation for this halo emission: inverse Compton by
cosmic ray electrons distributed on larger scales than those commonly discussed
and with anomalously hard energy spectrum. 

I find it intriguing that the angular distribution of the halo emission
resembles that expected from pair annihilation of dark matter WIMPs in the
galactic halo, and moreover, that the gamma-ray intensity is similar to that
expected from annihilations of a thermal relic with present mass density
0.1--0.2 of the critical density. Namely, the emission at $ b \ge 20^\circ $ is
approximately constant at a given angular distance from the galactic center --
except for a region around $(b,l) = (60^\circ,45^\circ) $, correlated to the
position of the Moon, and a region around $(b,l) = (190^\circ, -30^\circ) $,
where there is a local cloud ($b$ and $l$ are the galactic latitude and
longitude, respectively). Dixon et al.~\cite{Dixon98} argue against the
possibility of WIMP annihilations on the base that direct annihilation of
neutralinos into photons would give too low a gamma-ray signal. However most of
the photons from WIMP annihilations are usually not produced directly but come
from the decay of neutral pions generated in the particle cascades following
annihilation.

In this letter I show that WIMP annihilation can account for both the intensity
and the spatial distribution of the halo emission. A model independent
analysis, although instructive (see Ref.~\cite{Ringberg}), would necessarily be
vague. So to be concrete, I introduce a particle physics model with a suitable
dark matter candidate that satisfies present observational and experimental
constraints.

Consider a few GeV Majorana fermion in a model with an extended Higgs sector.
The Higgs sector contains two Higgs doublets $H_1$ and $H_2$ and a Higgs
singlet $N$. Let the Higgs potential be
\begin{eqnarray}
\label{eq:V}
  V_{\rm Higgs} = && 
  \lambda_1 \left( H_1^\dagger H_1 - v_1^2 \right)^2 + 
  \lambda_2 \left( H_2^\dagger H_2 - v_2^2 \right)^2 + \nonumber\\&&
  \lambda_3 \left| N^2 - v_N^2 \right|^2 + 
  \lambda_4 \left| H_1 H_2 - v_1 v_2 \right|^2 + \\&&
  \lambda_5 \left| H_1 H_2 - v_1 v_2 + N^2 - v_N^2 \right|^2 + 
  \lambda_6 \left| H_1^\dagger H_2 \right|^2 . \nonumber
\end{eqnarray}
Taking all the $\lambda$'s real and positive guarantees that the absolute
minimum of the potential is at $ \langle H_1^0 \rangle = v_1$, $ \langle
H_2^0 \rangle = v_2 $, and $ \langle N \rangle = v_N $, with zero vacuum
expectation values for all the other fields.

There are 6 physical Higgs bosons: one charged Higgs boson $H^\pm$, two neutral
``pseudoscalars'' $P_1$ and $P_2$, and three neutral ``scalars'' $S_1$, $S_2$,
and $S_3$ (I label the $P$s and $S$s in order of increasing mass). The charged
Higgs boson is $ H^{+} = H^{-*} \sin\beta + H_2^{+} \cos\beta $, and its mass
is $ m_{H^+} = \lambda_6^{1/2} v $, where $\tan\beta=v_2/v_1$ and $v =
\sqrt{v_1^2 + v_2^2} $.

The two pseudoscalars are linear combinations $ P_k = U^P_{k1} A + U^P_{k2}
N_I$ of $ A = \sqrt{2} ( \sin\beta \mathop{\rm Im} H_1^0 + \cos\beta
\mathop{\rm Im} H_2^0 ) $ and $ N_I = \sqrt{2} \mathop{\rm Im} N$. The unitary
matrix $U^P$ diagonalizes
\begin{equation}
  {\cal M}_P^2 = 
  \left[ \begin{array}{cc}
    (\lambda_4 + \lambda_5) v^2 &
     2 \lambda_5 v v_N \\
    2 \lambda_5 v v_N &
     4 (\lambda_3 + \lambda_4) v_N^2
  \end{array} \right] .
\end{equation}
The three scalars are linear combinations $ S_j = U^S_{j1} H_{1R} + U^S_{j2}
H_{2R} + U^S_{j3} N_R$ of $H_{1R} = \sqrt{2} \mathop{\rm Re} H_1^0$,
$H_{2R}=\sqrt{2} \mathop{\rm Re} H_2^0$, and $N_R=\sqrt{2} \mathop{\rm Re} N$.
The unitary matrix $U^S$ diagonalizes
\begin{equation}
 {\cal M}_S^2 = 
  \left[ \begin{array}{ccc}
    4 \lambda_1 v_1^2 + \lambda_{45} v_2^2 &
     \lambda_{45} v_1 v_2 & 
      2 \lambda_5 v_2 v_N \\
    \lambda_{45} v_1 v_2 &
     4 \lambda_2 v_2^2 + \lambda_{45} v_1^2 &
      2 \lambda_5 v_1 v_N \\
    2 \lambda_5 v_2 v_N &
     2 \lambda_5 v_1 v_N &
      4 \lambda_{35} v_N^2
  \end{array} \right] , 
\end{equation}
where $\lambda_{ij}=\lambda_i+\lambda_j$.

The Higgs bosons couple to the ordinary leptons and quarks and to the new
Majorana fermion $\chi$ through the Yukawa terms
\begin{equation}
  {\cal L}_{\rm Yukawa} = h_d \overline{Q} H_1 d_R + h_u \overline{Q} H_2 u_R +
  h_l \overline{L} H_1 e_R + h_\chi \overline{\chi} \chi N .
\end{equation}
Here $Q=(u_L,d_L)$ denotes the SU(2) quark doublets, $L=(\nu_L,e_L)$ the SU(2)
lepton doublets, and $d_R,u_R,e_R$ the SU(2) singlets. Generation indices are
suppressed.

When the Higgs fields get vacuum expectation values, the up--type quarks, the
down--type quarks and the charged leptons acquire masses $m_u = h_u v_2$, $m_d
= h_d v_1$ and $m_l = h_l v_1$, and the new Majorana fermion acquires mass $
m_\chi = h v_N$.  The interaction terms of the Higgs fields with the quarks,
the leptons and the $\chi$ read:
\begin{eqnarray}
  {\cal L}_{\rm int} &=& 
\frac{g m_{\chi}}{2m_W} \frac{v}{v_N} \left[ 
  \overline{\chi} \chi  S_j U^S_{j3} + 
  i \overline{\chi} \gamma_5 \chi P_k U^P_{k2} 
\right] \nonumber \\ &+&
\frac{g m_u}{2m_W} \left[ 
  \overline{u} u  S_j \frac{U^S_{j2}} {\sin\beta}  + 
  i \overline{u} \gamma_5 u P_k  U^P_{k1} \cot\beta 
\right] \nonumber \\ &+&
\frac{g m_d}{2m_W} \left[ 
  \overline{d} d S_j \frac{U^S_{j1}} {\cos\beta} + 
  i \overline{d} \gamma_5 d P_k U^P_{k1} \tan\beta 
\right] .
\end{eqnarray}

Now I find the interesting region in parameter space in which the calculated
gamma-ray emission matches the Dixon et al.\ maps while satisfying present
observational and experimental constraints.

The gamma-ray intensity from $\chi\chi$ annihilations in the
galactic halo is given by
\begin{equation}
\label{phigamma}
\phi_{\gamma}(b,l,E) = n_{\gamma}(E) \,
\frac{ \sigma v }{ 4 \pi m_\chi^2 } \, \int \rho_\chi^2 d l ,
\end{equation}
where $ \phi_{\gamma}(b,l,E) $ is in photons/(cm$^2$~s~sr), $n_{\gamma}(E)$ is
the number of photons per annihilation with energy above $E$, $ \sigma v$ is
the $\chi\chi$ annihilation cross section times relative velocity at $v=0$, $
\rho_\chi $ is the $\chi$ mass density in the halo, and the integral is along
the line of sight. For a canonical halo, $ \rho(r) = \rho_{\rm loc} ( r_c^2 +
R^2 )/( r_c^2 + r^2 ) $, the integral depends only on the angular distance
$\psi$ from the galactic center, and is practically independent of the halo
core radius $r_c$ at $\psi=56^\circ$, where it is $\int \rho_\chi^2 dl = 2.8
\rho_{\rm loc} R$.  Notice that for this halo model, the intensities at
$\psi=40^\circ$ and at $\psi=60^\circ$ are approximately in the ratio 2:1 as on
the Dixon et al.\ maps.

I consider interesting those parameter values for which the calculated
gamma-ray flux at $\psi=56^\circ$ is between 7 and 8 times 10$^{-7}$
photons/cm$^2$/s/sr for a local dark matter density $\rho_{\rm loc}$ in the
range 0.3--0.5 GeV/cm$^3$ and a distance from the galactic center $R=8$ kpc.
Fig.~1 shows the interesting region in the $\sigma v$--$m_\chi$ plane (a point
for each choice of parameter values). The figure includes the bounds discussed
in the following.

\begin{figure}[tbp]
\label{figmsv}
\epsfxsize = 0.8\hsize \epsfbox{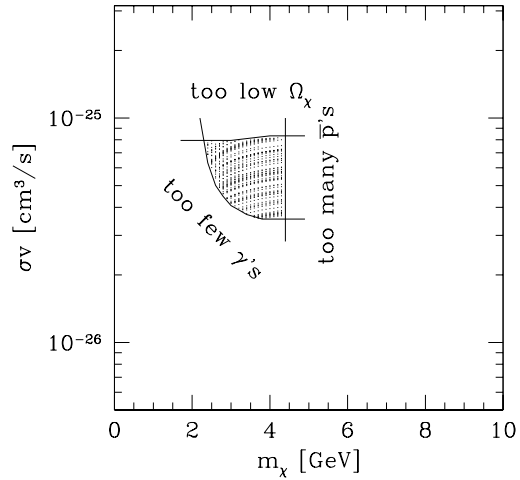}
\caption{Region in the $ \sigma v $ -- $m_\chi$ plane where 
the $\chi$ particle can explain the gamma-ray halo signal and satisfy all
experimental constraints considered.} 
\end{figure}

The most important constraint on the interesting region comes from the measured
flux of cosmic ray antiprotons. If the gamma-rays are produced in jets
originated by quarks, there is an associated production of antiprotons.  The
ratio of antiproton and gamma-ray fluxes is independent of the WIMP
annihilation cross section and of the local mass density, and the relative
number of antiprotons and photons per annihilation is fixed by the physics of
jets to a number of order 1. Since the antiprotons are confined in the galaxy
for $\sim 10^8$ yr while the gamma-rays from the halo take no more than $\sim
10^5$ yr to reach us, the antiproton flux is $\sim 10^3$ times larger than the
gamma-ray flux.  More precisely, the antiproton flux at a ${\rm \bar{p}}$
kinetic energy $T$ is
\begin{equation}
  \phi_{\rm \bar{p}}(T) =  \frac{ d N_{\rm \bar{p}}}{ d T} \,
  \frac{ \sigma v }{ 4 \pi m_\chi^2 } \, \rho_{\rm
    loc}^2 v_{\rm \bar{p}} t_{\rm cont} \mu .
\end{equation}
For the antiproton spectrum per annihilation $d N_{\rm \bar{p}}/ d T$ I use the
Lund Monte-Carlo, and I evaluate the containment time $t_{\rm cont}$ in a
diffusion model~\cite{WebberChardonnetBottino}: $t_{\rm cont} \simeq (1+p/{\rm
  GeV})^{-0.6} 5 \times 10^{15} $ s.  The factor $ \mu =
[T(T+2m_{\rm\bar{p}})]/[(T+\Delta)(T+\Delta+2m_{\rm\bar{p}})]$ takes into
account solar modulation.  Given the uncertainties in the antiproton
propagation in the galaxy and in the effect of the solar modulation, I accept
models with $\phi_{\rm \bar{p}}(200{\rm MeV}) < 3 \times 10^{-6} $ ${\rm
  \bar{p}}$/cm$^2$/s/sr/GeV~\cite{BESS95} (taking $\Delta=600$ MeV). This bound
effectively limits the viable decay channels to $\chi\chi\to\tau^+\tau^-$.

The $\chi$ relic density can be obtained with a standard
procedure~\cite{KolbTurnerGondolo91},
\begin{equation}
\Omega_\chi h^2 = \frac{ 1.023 \times 10^{-27} {\rm cm^3/s}  }{
  \int_0^{x_f} \langle \sigma v \rangle g_{*}^{1/2} d x } ,
\end{equation}
where $x_f$ is the solution of $ x_f^{-1} + {1\over2} \ln (g_{*} / x_f) = 80.4
+ \ln( m_\chi \langle \sigma v \rangle ) $ with $m_\chi$ in GeV and $\langle
\sigma v \rangle $ in cm$^3$/s. Here $ \langle \sigma v \rangle $ is the
thermally averaged annihilation cross section at temperature $xm$. Since
freeze-out occurs below the QCD phase transition, I take $g_{*} = 81$ for the
relativistic degrees of freedom at freeze-out. The result of the relic density
calculation is shown in Fig.~2.  The relic abundance in the interesting region
turns out to be in the cosmologically interesting range: $\Omega_\chi \sim 0.1$
for a Hubble constant of 60 km/s/Mpc.

\begin{figure}[tbp]
\label{figmo}
\epsfxsize = 0.8\hsize \epsfbox{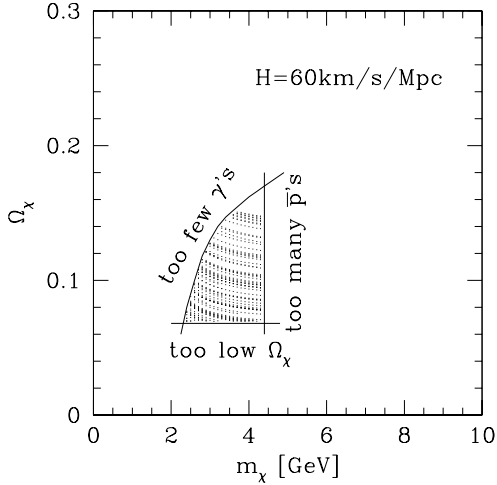}
\caption{The $\chi$ relic density for $\chi$ particles that 
can explain the gamma-ray halo signal and satisfy all
experimental constraints considered.}
\end{figure}

In the interesting region, two Higgs bosons, $S_1$ and $P_1$, are light, with
masses between 2 and 16 GeV. They are mostly singlets and their production in
colliders is very suppressed. First, I consider the LEP bounds on the search of
Higgs bosons in two-doublet models.  Since there is an additional Higgs
singlet, $\sin^2(\beta-\alpha)$ in the bound from $e^+ e^- \to Z h$ must be
replaced with $| U^S_{11} \cos\beta + U^S_{12} \sin \beta |^2$, and
$\cos^2(\beta-\alpha)$ in the bound from $e^+ e^- \to hA$ must be replaced with
$ | U^P_{12} U^S_{12} - U^P_{11} U^S_{11} |^2$. In the allowed region, the
reinterpreted $\sin^2(\beta-\alpha)$ and $\cos^2(\beta-\alpha)$ are smaller
than $5 \times 10^{-5}$, and are not excluded by accelerator
searches~\cite{Decamp92}.  Secondly, I consider Higgs bremsstrahlung from final
state leptons and quarks in Z decays. I find
\begin{equation}
\label{brems}
\frac{ \Gamma\!(Z\to f\overline{f}X) }{ \Gamma\!(Z\to f\overline{f}) }
= \frac{ \sqrt{2} G_F } { 4 \pi^2} \, g\!\left(\frac{m_X}{m_Z}\right) \,
c_f \, m_f^2 A_{Xf}^2 ,
\end{equation}
where $A_{P_if} = U^P_{i1} \cot\beta$, $A_{S_if} = U^S_{i1}/\cos\beta$ for
up--type quarks, $A_{P_if} = U^P_{i1} \tan\beta$, $A_{S_if} =
U^S_{i2}/\sin\beta$ for down--type quarks and leptons, $c_f = 3$ for quarks and
$c_f=1$ for leptons. The function $g(y)$ comes from the phase-space
integration, and is $g(y) \cong 1$ at $m_X=12$ GeV. The most stringent
constraint from the LEP results on two leptons + two jets
production~\cite{Decamp92} is $ \Gamma\!(Z\to f\overline{f}X) / \Gamma\!(Z\to
f\overline{f}) < 1.5 \times 10^{-4} $ at $m_X = 12 $ GeV.  The highest values I
find in the interesting region of my model are of order $10^{-6}$, and so this
constraint is easily satisfied.

In addition to accelerator bounds, the $\chi$ particle must satisfy present
bounds from direct and indirect dark matter searches. The cross section for
elastic $\chi$--proton scattering, which is spin-independent, is
\begin{equation}
  \sigma_{\chi p} = \frac{G_F^2}{\pi} \, \frac{2 m_p^4 m_\chi^4}{\left( m_p +
    m_\chi \right)^2} \, \frac{v^2}{v_N^2} \, \left[ \sum_{j=1}^3
  \frac{U^S_{j3}}{m_{S_j}^2} \left( \frac{k_d U^S_{j1}}{\cos\beta} + \frac{ k_u
    U^S_{j2}}{\sin\beta} \right) \right]^2 ,
\end{equation}
where $ k_d = \langle m_d \overline{d} d + m_s \overline{s} s + m_b
\overline{b} b \rangle = 0.21$ and $ k_u = \langle m_u \overline{u} u + m_c
\overline{c} c + m_t \overline{t} t \rangle = 0.15$. Fig.~3 shows a comparison
of this cross section with the presently most stringent upper bound on few GeV
WIMPs~\cite{Bernabei98}. The bound touches the interesting region, but to
explore it all major improvements in sensitivity are needed, especially at low
thresholds~\cite{CRESST}.

\begin{figure}[tbp]
\label{figmsp}
\epsfxsize = 0.8\hsize \epsfbox{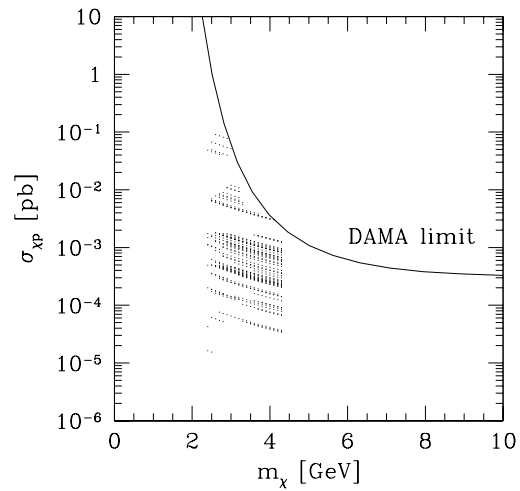}
\caption{The $\chi$--proton spin-independent scattering cross section versus
  the $\chi$ mass. The solid line is the present bound 
  from ref.~[7]; the dots show the interesting region.}
\end{figure}

The $\chi$ fermions can also accumulate in the Sun and in the Earth cores,
annihilate therein and produce GeV neutrinos.  Accumulation in the Earth is not
efficient because the $\chi$ mass is below the evaporation mass from the Earth
($\sim 12$ GeV), and no neutrino signal is expected from there. The evaporation
mass in the Sun ($\sim 3$ GeV) is in the middle of the interesting region.
Following Ref.~\cite{Griest87Gould87GGR91Kamionk95}, I have calculated the
neutrino flux from the Sun including evaporation, annihilation and capture, and
then the rate of contained events and the flux of through-going muons in
neutrino telescopes. I have imposed the experimental bounds in
Ref.~\cite{KamiokandeSuvorova97Montaruli98}.  Fig.~4 shows that, although the
present limits on through-going muons constrain the allowed region, it will be
difficult to probe models with $m_\chi$ below the evaporation mass of $\sim 3$
GeV because of the exponential suppression of the signal.

\begin{figure}[tbp]
\label{figmthru}
\epsfxsize = 0.8\hsize \epsfbox{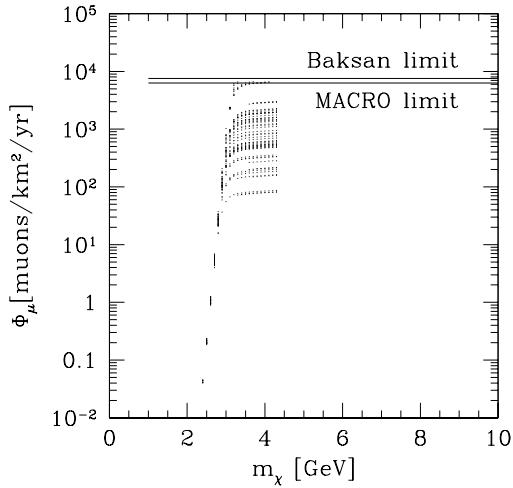}
\caption{The flux of through-going muons from the Sun induced by neutrinos from
  $\chi$ annihilations versus the $\chi$ mass. The solid lines are the Baksan
  and MACRO limits [10]; the dots show the interesting region.}
\end{figure}

The clearest test of WIMP annihilations in the halo is the shape of the
gamma-ray spectrum. It must show a sharp cutoff and a gamma-ray line at the
WIMP mass. The continuum spectrum is determined from jet physics. The gamma-ray
line intensity is proportional to the annihilation cross section into two
photons, which in the specific model presented here results in the range
$10^{-29}$--$10^{-30}$ cm$^3$/s (using formulas in Ref.~\cite{Bern}). The poor
energy resolution of present gamma-ray detectors unfortunately smears the line
out: even a tantalizing energy resolution of 0.1\% merely gives $10^{-12}$
photons/cm$^2$/s/sr/MeV in this model. Fig.~5 illustrates the gamma-ray
spectrum expected from $\chi\chi$ annihilations for $\chi$ particles in the
interesting region. For comparison, the figure indicates the measured
extragalactic flux~\cite{extragal}. Since the antiproton constraint limits the
$\chi$ mass to less than 4.3 GeV, a clear detection of halo gamma-rays
more energetic than this would be an indication against the present WIMP model.

In conclusion, it is possible to explain the Dixon et al.\ gamma-ray halo as
due to WIMP annihilations in the galactic halo. As illustrated in an explicit
model, not only the intensity and spatial pattern of the halo emission can be
matched but also the relic density of the candidate WIMP can be in the
cosmologically interesting domain. The model is quite predictive and can be
tested with a variety of techniques, above all by accurately measuring
the energy spectrum of the halo gamma-ray emission.

{~}

I thank D.\ D.\ Dixon for discussions.

\begin{figure}[tbp]
\label{figgs}
\epsfxsize = 0.8\hsize \epsfbox{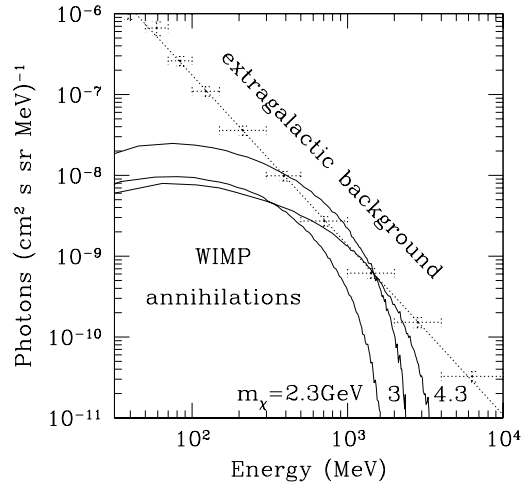}
\caption{The predicted gamma-ray spectrum from WIMP annihilations in the halo
  compared with the measured extragalactic flux~[12].  Solid lines correspond
  to $\chi$ masses of 2.3, 3, and 4.3 GeV, and they delineate the range of
  expected signals.}
\end{figure}

\end{document}